**Monitoring of water sorption and swelling of potato starch-glycerol extruded blend by magnetic resonance imaging and multivariate curve resolution.**

Silvia Mas Garcia[1,2*], Jean-Michel Roger[1,2], Ruzica Ferbus[3], Denis Lourdin[4], Corinne Rondeau-Mouro[3]

[1] ITAP, INRAE, Institut Agro, University Montpellier, 34196 Montpellier, France

[2] ChemHouse Research Group, 34196 Montpellier, France

[3] INRAE, UR1466 OPAALE, 17 Avenue de Cucillé, CS 64427, F-35044 Rennes, France

[4] INRAE, UR1268 Biopolymères Interactions Assemblages, 44300 Nantes, France

[*]Author for correspondence:

E-mail: silvia.mas-garcia@inrae.fr

**Abstract**

Magnetic resonance microimaging (MRμI) is an outstanding technique for studying water transfers in millimetric bio-based materials in a non-destructive and non-invasive manner. However, depending on the composition of the material, monitoring and quantification of these transfers can be very complex, and hence reliable image processing and analysis tools are necessary. In this study, a combination of MRμI and multivariate curve resolution-alternating least squares (MCR-ALS) is proposed to monitor the water ingress into a potato starch extruded blend containing 20% glycerol that was shown to have interesting properties for biomedical, textile, and food applications.

In this work, the main purpose of MCR is to provide spectral signatures and distribution maps of the components involved in the water uptake process that occurs over time with various kinetics. This approach allowed the description of the system evolution at a global (image) and a local (pixel) level, hence, permitted the resolution of two waterfronts, at two different times into the blend that could not be resolved by any other mathematical processing method usually used in magnetic resonance imaging (MRI). The results were supplemented by scanning electron microscopy (SEM) observations in order to interpret these two waterfronts in a biological and physico-chemical point of view.

**Keywords:** Multivariate curve resolution-alternating least squares (MCR-ALS), swelling process, potato starch-glycerol extruded blend, magnetic resonance microimaging (MRμI).

# 1. Introduction

Over the past decade, there has been a growing interest in using starch in diverse applications other than food, especially in the medical and pharmaceutical fields because of its biodegradability, non-toxicity, and overwhelmed abundance [1–4]. Starches can be obtained from many different sources (such as maize, wheat, potatoes, ...) and offer promising natural polymers due to their low material cost and ability to be processed with conventional plastic processing equipment[5]. They have received extensive attention in relation to their unique structural and physicochemical properties. Among them, water sorption and swelling power are of great interest because of their relevance to the storage, drying, enzymatic activities, and processing of starch-based materials, but also for the formulation of controlled release materials[6].

Water dynamics in such materials are analysed by monitoring their affected physical or chemical properties that show time-dependent changes. Among the different techniques approaching water mobility, magnetic resonance imaging (MRI) is based on the magnetic properties of atomic nuclei and in particular hydrogen. It gives two- or three-dimensional images of the interior of the object in a non-invasive manner with relatively high contrast sensitivity. For sample dimension of few centimeters, magnetic resonance microimaging (MRµI) probes are used in high-field magnets giving sufficient signal-to-noise ratio and an in-plane resolution of up to a few tens of µm². Because of its sensitivity to the number of hydrogen nuclei, MRµI makes it possible to discriminate between the different types of water mobility, an advantage that is found in certain materials containing enough hydrogen nuclei known as «mobile». This notion of mobility applies to both atoms and molecules themselves. Also called molecular dynamic phenomena, they are by nature very complex and have a random and statistical character. Reorientation, translation movements as well as transfers or flows of molecules in space, occur at different scales and can be apprehended using MRµI. Using this technique , the NMR signal can be measured in each element of imaged volume (voxel, equivalent to pixel with a thickness), at a particular moment or over time. The signal intensity is determined by four basic parameters, proton density, T1, T2 relaxation time, or flow. The continuous monitoring of water transfers in biological materials, at each voxel, using multiple T2-, T1- or Diffusion-weighted images were shown very useful to get information on the microscopic water distribution in bio-based materials, in complement to the dynamic structural changes of the same materials [7–9]. Like in time domain –NMR (TD-NMR), the measured relaxation or diffusion signal in each voxel can be modelled by an exponential decay. In case of a multi-exponential behavior, extracting the multi-exponential parameters at each voxel is a challenge. Usually in MRµI, mathematical models of molecular dynamic phenomena use parametric representation based on a finite number of components inside each voxel. In Multiple Spin Echo (Multi-SE) acquisitions, given the weak number of consecutive equally spaced echoes used, the T2-weighted images are obtained with a mono-exponential decaying model. However, such model requires the solving of a nonlinear estimation problem[10]. For that, the most classical approach is the discrete exponential fitting of data. This approach obviously relies on an appropriate validation procedure for determining the correct number of components in the signals, to avoid overfitting. In case, the number of signal components is not known and/or to avoid the problem of algorithm initialization, the use of other methods for estimating relaxation times may be more relevant. These methods are generally based on the inverse Laplace transform [11]. While no initial condition is necessary in this case, as for any equation to partial derivatives, it is generally necessary to specify boundary conditions that often makes the problem mathematically «ill-posed» with multiple solutions. Indeed, the presence of even small noise in the data affects the number of solutions and can make the data adjustment from problematic to impossible. However, for a multi-exponential model, it is necessary to increase the signal-to-noise ratio (SNR), which can be done by averaging the signal for each echo time on a homogeneous region of interest (ROI) of the images. However, this approach reduces spatial information to the scale of a ROI.

To deal with these limitations, this work aims to propose the use of chemometrics methods as an alternative. Chemometrics approaches allow extracting useful signals from very cluttered and noisy data[12,13]. These methods allow the processing of curve shaped data, such as spectra[14]. From a collection of spectra, these methods allow the identification of a limited number of spectral signatures (loadings) underlying the studied system. Scores then represent each sample, projected on these loadings. Linear algebra tools, which perform linear combination of the spectra of the original collection, perform the extraction of this information and, hence, the extracted loadings have a very good signal-to-noise ratio. Among chemometrics tools, the multi curve resolution approaches[15] carry out the loadings/scores decomposition under specific constraints. Giving constraints related to chemical laws, as loadings positivity and the unitary sum of scores, allows the MCR to yield pure spectra as loadings and concentrations as scores. When MCR is applied to TD-RMN spectra, the pure spectra are NMR signals that can be processed by an inverse Laplace transform without numerical instability problems. This provides reliable information on the compounds present in the system[16]. Moreover, concentration profiles estimated by MCR allow highlighting temporal phenomena, if the system monitored is process-related, and/or increase the knowledge of spatial structures if the raw data are organised in images.

The novelty and main goal of this work is showing the potential of a MCR method, named MCR-alternating least squares (MCR-ALS)[17,18], to analyze a series of 2D images (constituted of pixels) collected from a slice of 3D MR images acquired during a swelling process. From the resolved spectra, after an inverse Laplace transformation, identification of the water transfer forms can be achieved. The process evolution is described by the resolved distribution maps, which provide information at global and local (pixel) levels. The advantages of monitoring this kind of process by MRµI using chemometrics instead of classical processing methods will be shown through the monitoring of a potato starch blend swelling process. An interpretation of hydration mechanism will be given on the basis of scanning electron microscopy experiments.

## 2. Material and Methods

### 2.1 Sample

The sample consisted in a starch-glycerol blend immersed in deuterated water. It was composed of potato starch purchased from Roquette (Lestrem, France) and glycerol (Merck, purity > 98 %) at a ratio of 20 % in wet basis. The preparation of the starch-glycerol blend were reported in Chevigny et al[19]. Briefly, a mixture of starch powder, glycerol and water was introduced in a single screw extruder equipped with a die outlet diameter of 3mm. The powder was melted during the thermomechanical treatment realised at 115 °C and a flow rate of 3.1 g.min$^{-1}$ allowed a specific mechanical energy of 122 J.g$^{-1}$ applied to the material. At the exit of the die, the starch-glycerol blend was a continuous and transparent cylinder of 3.5 mm of diameter.

### 2.2 Magnetic resonance microimaging (MRµI)

The MRµI setup were described in Chevigny et al[19]. The ParaVision software (PV6, Bruker, France) was used for the acquisition of data.

A matrix of 128 × 128 pixels was used with a field of view (FOV) of 10mm × 10mm and a slice thickness of 500 µm. The image resolution was 78.12 µm × 78.12 µm in plan. A multi slice multi echo (MSME) sequence was acquired with first echo time (TE1) of 5 ms, 5 averages, an inter-echo spacing ΔTE of 5 ms and 32 echoes per echo train. A repetition time (TR) of 1.5 s was chosen leading to an acquisition time of about 15min for each image, respectively.

### 2.3 Scanning electron microscopy (SEM)

SEM observations were performed on a field emission gun scanning electron microscope (Thermo Fischer Scientific, Quattro S, Waltham, Massachusetts, USA). Extruded starch cylinder was immersed in water during a controlled time, afterwards samples were cut in order to obtain discs of approximately 1mm thick and immediately placed on the ESEM sample stage. Samples surface were observed under ESEM mode at a pressure of 900Pa in the chamber of the microscope and a temperature of 4°C. Images were recorded at an acceleration voltage of 15kV and a spot size of 3.

**2.4 Chemometrics**

In MRµI, the generated data consists of a three-dimensional array known as a hyperspectral cube, the x- and y-axis correspond to the pixel coordinates and the z-axis corresponds to the echo times registered in each pixel. In this work, Multivariate Curve Resolution-Alternating Least Squares (MCR-ALS) were used to exploit the chemical information of the NMR signal in each pixel of the image, providing a spatial distribution of the chemical composition of the sample. Beforehand, the background pixels were masked out and then the hyperspectral cube was unfolded a **D** matrix to be analysed.

**2.4.1 Multivariate Curve Resolution- Alternating Least Squares (MCR-ALS)**

MCR-ALS is based on a bilinear decomposition of data by means of an alternating least squares (ALS) algorithm, that can be described in linear algebra terms by:

$$\mathbf{D} = \mathbf{CS}^T + \mathbf{E} \qquad \text{Equation 1}$$

where **D** is the experimental matrix that contains the spectra of all the pixels in the image, **C** is the matrix of the concentration profiles of the constituents present in NMR signal of the image, $\mathbf{S}^T$ the matrix of their related pure spectra and **E** is the matrix expressing the error or variance unexplained by the bilinear model[17,18]. Each column of the resolved **C** matrix can be refolded to recover the original 2D spatial image structure and then pure distribution maps are obtained (See Figure 1a).

MCR-ALS is an iterative method that requires both the input of the number of pure components and an initial estimate of their spectral signature ($S^T$). In this work, singular value decomposition (SVD)[20] approach was used to estimate the number of components. MCR-ALS models with different components should be tested; each added component must clearly increase the explained variance in the data and lead to a model with interpretable concentration and spectral profiles. Initial estimates were generated using simple-to-use interactive self-modelling mixture analysis (SIMPLISMA) method[21].

MCR-ALS analysis could converge to different sets of concentrations and spectral profiles that equality fit the data. This ambiguity is usually suppressed or decreased by introducing certain constraints[17,22] commonly related to prior knowledge about the problem faced, so that it is possible to obtain easier-to-interpret solutions. In this work, non-negative concentrations and non-negative spectral intensities were assumed because neither the concentration nor the spectral intensity of a chemical component should be negative at any pixel. Normalization was applied to $S^T$ matrix during the MCR-ALS iterative process in order to obtain concentrations $C$ that can be directly compared between them. Besides, additional constraint of form was also applied in some profiles within $S^T$ imposing an exponential decay which is the shape expected for MRµI spectral signatures.

In a process monitoring, the simultaneous analysis of all recorded images is mandatory to obtain a complete and reliable description of the process under study and less affected by ambiguity phenomenon [17,18]. In this case, multiset structures were built that contain different submatrices $D_i$, linked to each of the individual images acquired during the hydration process, arranged one on top of each other to build a column-wise augmented matrix $\mathbf{D_{aug}}$. The bilinear model in Equation 1 is now extended to the augmented data set as shown in Equation 2:

$$\mathbf{D_{aug}}=[\mathbf{D_1};\mathbf{D_2};...;\mathbf{D_n}]=[\mathbf{C_1};\mathbf{C_2};...;\mathbf{C_n}]\mathbf{S^T}+[\mathbf{E_1};\mathbf{E_2};...;\mathbf{E_n}]=\mathbf{C_{aug}}\mathbf{S^T}+\mathbf{E_{aug}} \qquad \text{Equation 2}$$

where $\mathbf{C_{aug}}$ is a column-wise augmented matrix formed by as many submatrices $\mathbf{C_i}$ as images in the multiset, and $\mathbf{S^T}$ is a single data matrix, assumed to be common for all the images in the multiset. Thus, it is expected to extract more information about the general swelling behaviour of the analysed sample. The concentration profiles in each of these submatrices can be also refolded conveniently to recover the related distribution maps of each image. Kinetic profiles of the compounds involved in the swelling process can be derived by displaying the evolution of the intensities of the resolved distribution maps as a function of the process time (see Figure 1b).

MCR-ALS distribution maps ($\mathbf{C}$ matrix) and pure spectra ($\mathbf{S^T}$ matrix) are excellent low dimension, noise-filtered meaningful basis of the pixel and the spectral space of the image, which may be further processed to obtain additional information. In this work, inversion of the Laplace transform[23] was applied on the pure spectra in order to interpret the meaning of the resolved profiles and understand the water transfer during swelling process.

The application of MCR-ALS has been performed in MATLAB platform (Version 2015b, MathWorks Inc., Natick, MA, USA) using the MCR GUI (multivariate curve resolution graphical user interface) developed by the chemometrics group of Universitat de Barcelona and IDAEA-CSIC[24], which is can be downloaded from the MCR webpage http://www.mcrals.info/.

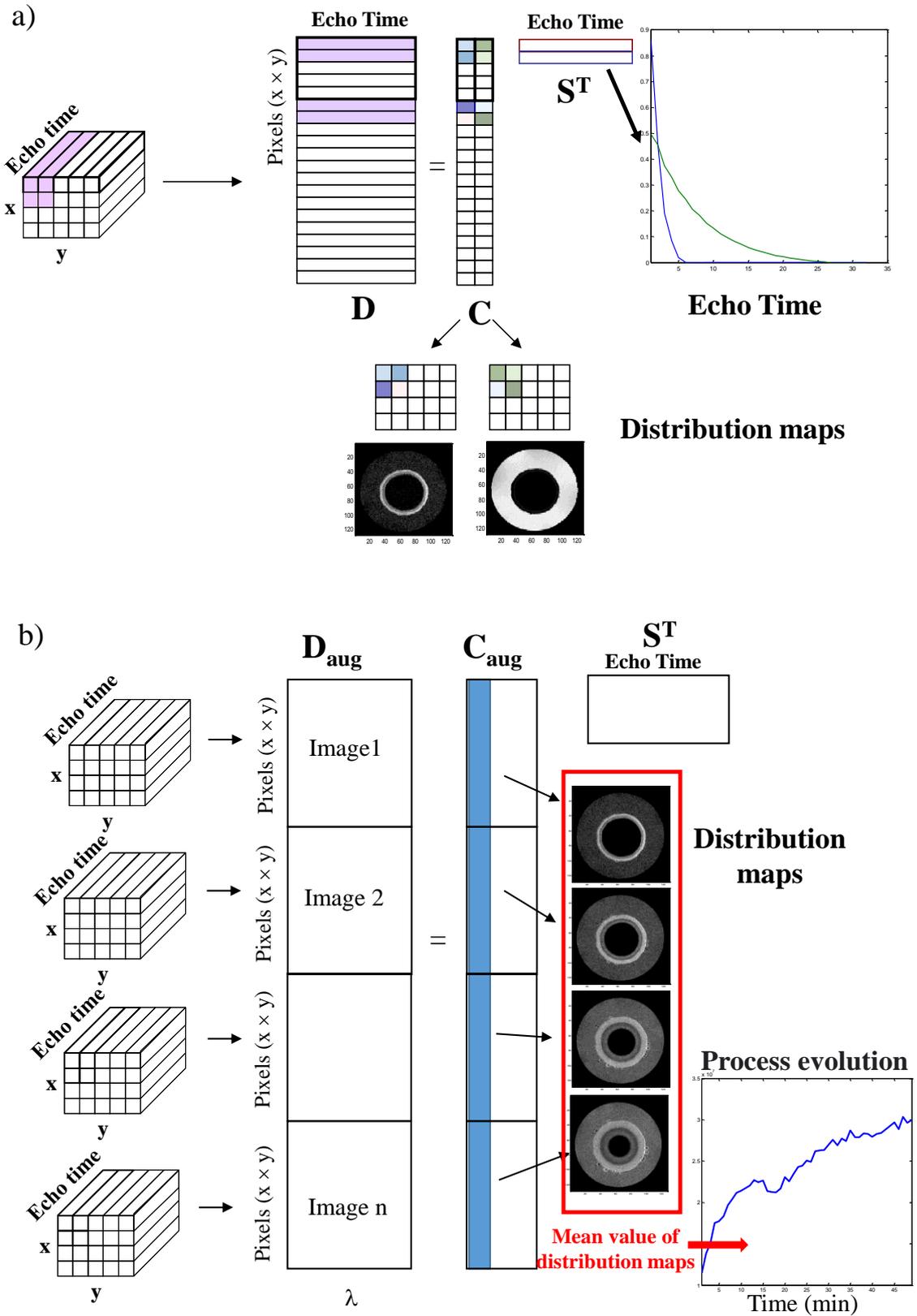

Figure 1. MCR application to a) an individual hyperspectral image, b) an image multiset structure.

## 3. Results and discussion

### 3.1 Raw data exploration

Figure 2 a-c shows the T2-weighted maps of the starch-glycerol blend at 0.78 H, 2.52 H and 21.30 H of imbibition and the total relaxing signal for each inter-echo spacing (ΔTE) recorded at the beginning, middle and end of the hydration process. The T2-weighted maps display a clear evolution of the swelling process. As shown in Chevigny et al. [19], this swelling is observed over time. It corresponds to an increase in diameter of 25.7 %, with an initial diameter of 4.68 ± 0.21 mm and a final one of 5.88 ± 0.01 mm, after 22 H of hydration. Very noisy raw spectra can be observed at the beginning, middle and end of the hydration process (Figure 2 d-f). This could make difficult the use of classical signal processing such as non negative least square fitting procedures and numerical inversions of ILT. Therefore, to find out a comprehensive description of the process by analyzing the whole images, the use of MCR-ALS was justified.

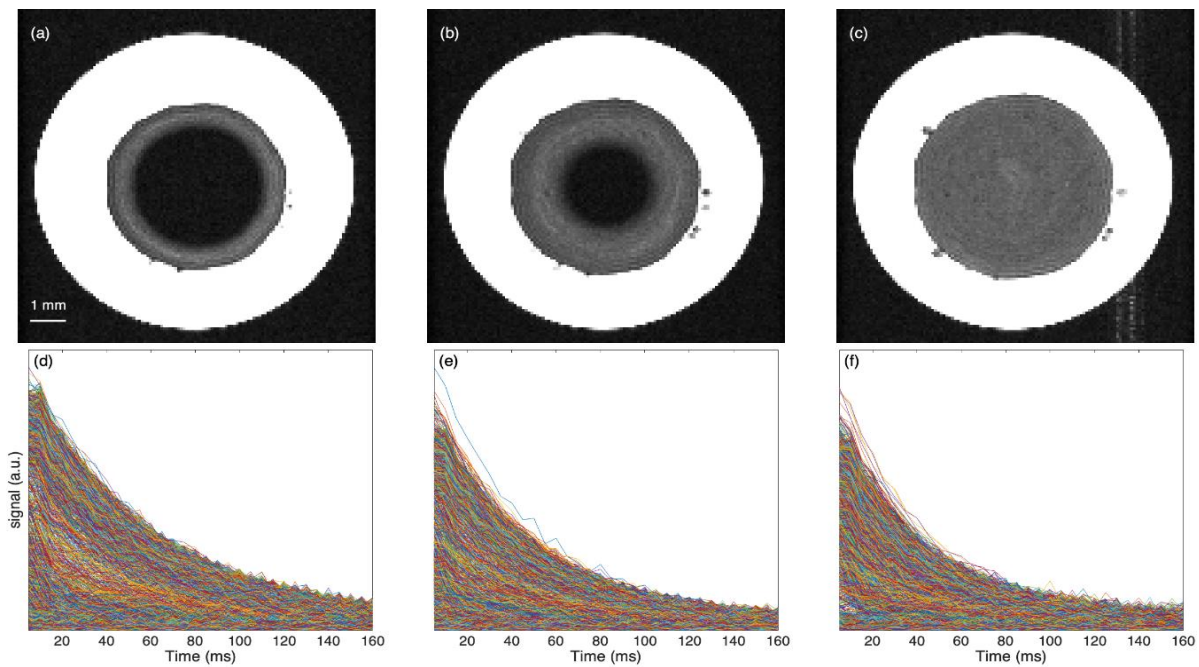

Figure 2. Raw global intensity maps (a-c) of the collected starch-glycerol blend images and their related raw spectra (d-f) recorded at 0.78 H (a,d), 2.52 H (b,e) and 21.30 H (c,f).

**3.2 MCR-ALS results**

This section shows the results of MCR-ALS analysis that have been carried out on the multiset structure formed by all the 49 images collected along the swelling monitoring of starch-glycerol blend at the different time process from 0.31 H to 21.30 H.

Firstly, an exploratory MCR-ALS analysis of the swelling evolution has been carried out applying the constraints of non-negativity in the spectral and concentration profiles and normalization in the spectral profiles. Resolution of four species was achieved with a satisfactory explained variance (99.8 %). The inclusion of a different number of species gave lower explained variance or unreliable spectra. From the shape of the spectra profiles (see figure S1 in supplementary information), it seems that one of the four resolved components, which does not present an expected decay shape, may not be involved in the process. The nature of this extra non-process contribution is unknown. However, it was needed to improve the resolution results. These extra contributions needed to perform the complete description of the dataset using MCR-ALS analysis were also observed in [25]. In this preliminary MCR-ALS analysis, the shape of all spectral signatures were not fully satisfactory because they do not present a real decay curve. Therefore, shape constraint was added to impose the decay curve shape. The variance explained obtained in the resolution using only non-negativity and normalization constraints was 99.8 % and now with the use of the spectral shape constraint is 97.7 %. The latter value indicates that the new constraint introduced a loss of fit of around 2 % but produced more meaningful spectral shapes and less ambiguous results. Once, meaningful and noised-filtered spectra are obtained, the application of the classical ILT could be carried on in order to interpret the meaning of the resolved profiles. Moreover, process evolution profiles can be obtained by plotting the mean values of the distribution maps of each constituent versus the process time as mentioned before in section 2.4.1.

## 3.3 Interpretation of results: T2 characterization and swelling

Figure 3a and b shows the MCR resolved spectra using the shape constraint and the results of applying ILT on them, respectively. Again, one of the four resolved components (the cyan spectra profile), not involved in the swelling process, was needed to achieve a complete description of the system. Considering this spectra profile as non-process contribution, ILT on this spectra profile was not applied. ILT of blue, red and green spectra gave each one relaxation time at 7, 17 and 33 ms, respectively (Figure 4b). These values were coherent with expected $T_2$ values[26] for doped water imbibing and being outside the starch-glycerol blend (33 ms in green), while two shorter T2 values originated from the water inside the imbibed starch-glycerol starch-glycerol blend (7 and 17 ms in blue and red, respectively). These two relaxation times could be assigned to two distinct processes of imbibition of the starch-glycerol blend.

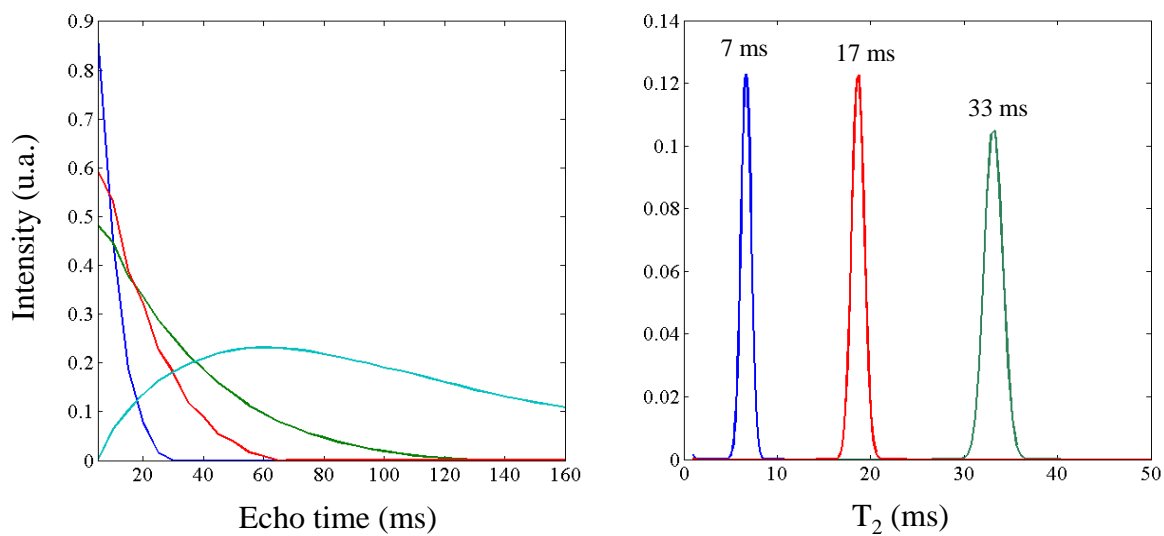

Figure 3. MCR-ALS multiset results applying spectral shape constraint: a) MCR resolved spectra of the species involved in the hydration of starch-glycerol blend; b) T2 relaxation curves

Figure 4 shows the distribution maps of the three process contributions (blue, red and green), which allow following the process evolution of species over time with a spatial resolution. To simplify, resolved distribution maps of only the images recorded at 0.78 H, 2.52 H and 21.30 H as in Figure 2 are shown. Distribution maps of the non-process contribution (cyan) was omitted because it does not provide physical/chemical meaningful information. From the distribution maps of the blue (C1), green (C2) and red (C3) contributions, two imbibition processes or kinetics could be observed. A thin layer at the edge of the sample is formed as soon as the water transfer starts to occur. Then, two zones can be distinguished in the C1 contribution, a water phase at the sample edges that expands over time and an inner layer of water, forming a kind of ring in the starch-glycerol blend. The observation of a hypersignal with a ring shape proves that water goes through the sample just before its strong interaction with the solid matrix. After 2.52 H, the second water wave becomes larger while the first and quicker wave comes closer to the center, due to the extension of the sample hydration. Afterwards, the appearance of a homogeneous grey signal in the whole blend and the disappearance of the noisy central zone (dry sample without any contact with water) evidences the complete hydration of the whole sample. This equilibrium appeared to be reached at 21.30 H of immersion. Regarding the C2 contribution, it is difficult to distinguish between these two water waves. Instead, a signal in the form of a very noisy ring that evolves over time in the starch-glycerol blend is observed until it disappears. This weak signal is difficult to interpret. On the other hand, the hypersignal observed outside the starch-glycerol blend represents water with relatively high mobility.

Finally, the C3 image shows the hypersignal composed of the sum of the rapid first and the slower second water waves. At the same time, there is an increase in the signal outside the starch-glycerol blend. It can be assumed that this increase is due to the glycerol and/or starch polymer fragments leaching out as the water diffused into the starch-glycerol blend[26]. C3 would come from the water inside the solid matrix and glycerol outside the blend.

Observing different transfer regimes may reflect the existence of cavities of heterogeneous sizes and shapes with the filling of larger cavities first and then smaller cavities. It can be assumed that these cavities are filled with glycerol before immersion and that the inflow of water into these cavities results in strong interactions with the surfaces that collapse by plasticizing the starch (the cavities close). This collapse is concomitant with the expulsion of glycerol outside the starch-glycerol blend.

The full set of distribution maps is provided as supplementary information under video format that displays better the evolution of the different compounds (see movie S1).

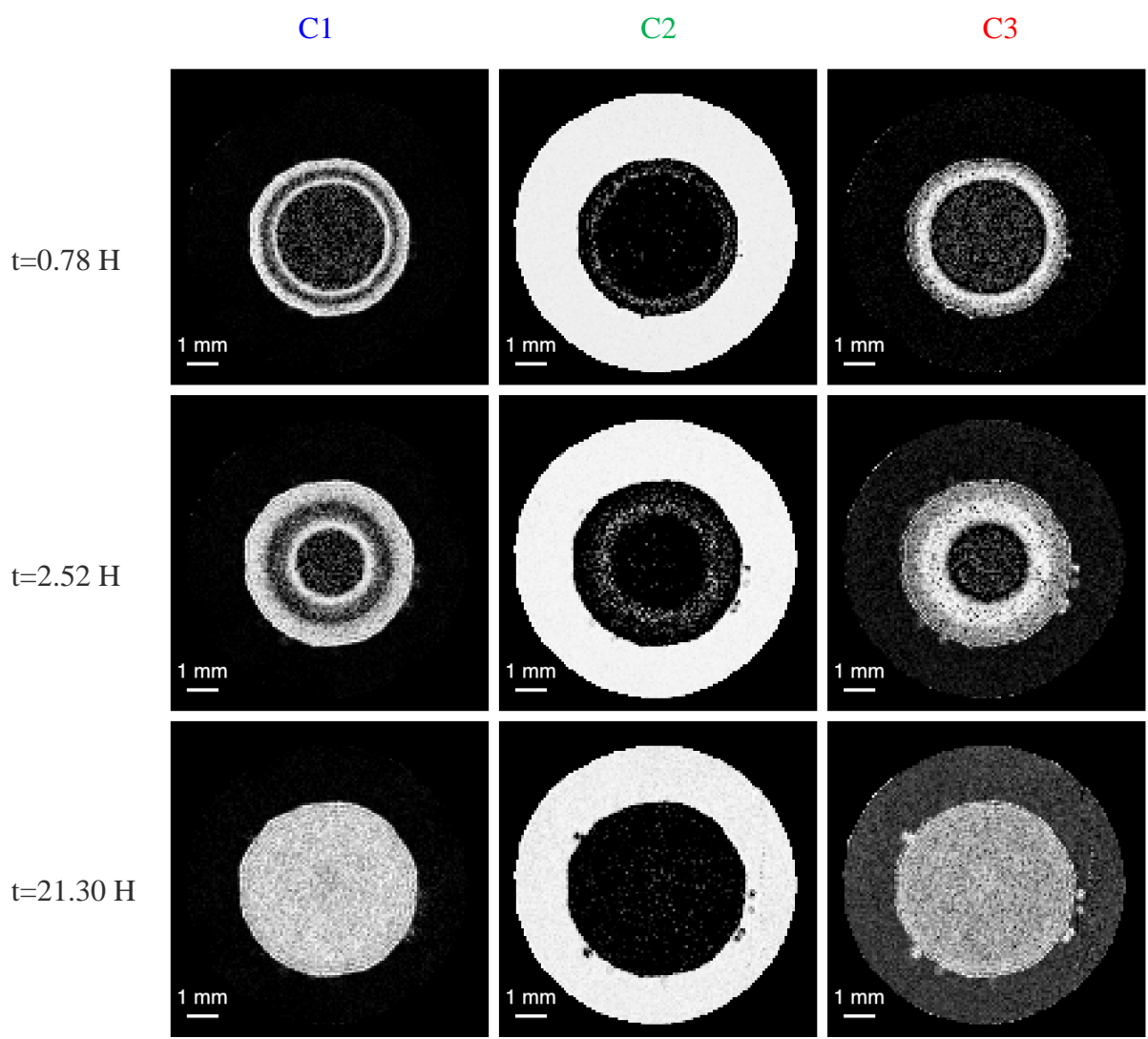

Figure 4. MCR distributions maps of the three species involved in the swelling process of starch-glycerol blend, at 0.78 H, 2.52 H, 21.30 H.

Global kinetic process profiles can be obtained by plotting the mean values of the distribution maps of each constituent versus the time process as mentioned before in section 2.4.1 (see Figure 5). As expected, the water diffusion and the water penetration front moving towards the center of the starch-glycerol blend resulted in an exponential increase of the imbibed starch-glycerol blend contributions (C1 and C3) and an exponential decrease of the external water contribution (C2). The kinetics of C2 and C3 (and at a less extent of C1), clearly indicate a more rapid evolution during the six first hours of imbibition and an afterwards stagnation. It is important to note that blue and red profiles, corresponding to imbibed starch-glycerol blend contributions present a similar kinetic evolution. However, differences in the spatial distribution and spectral signature allows distinguishing of these two contributions.

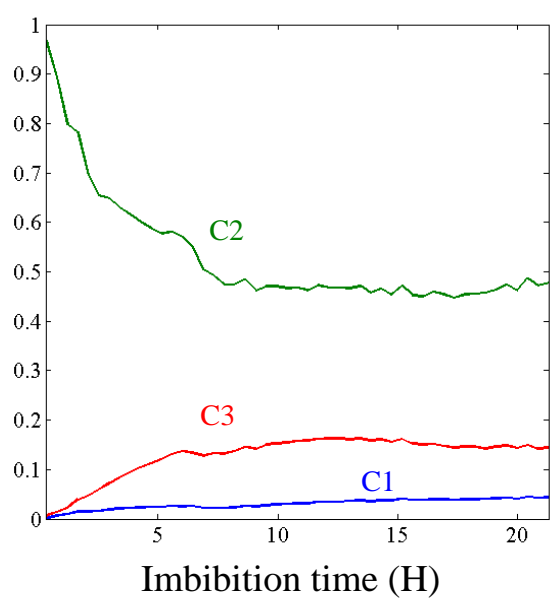

Figure 5. Representation of global temporal profiles of the three species involved in the swelling process.

It is also interesting to know how the process evolves at a local pixel level to find out whether there can be specificities unnoticed in global profiles, to explore a certain sample area or positions of particular interest. Figure 6 shows the concentration evolutions of C1, C2 and C3 at four distances from the center of the extruded sample. The first wave of water reaches the center of the sample (D= 0 mm) after 6 H of immersion. This very quick phenomenon results in the extinction of the C1 signal that reappears after 13 H of immersion in water but with a slower progression. This phenomenon is also observed in the C3 and C2 components that keep a certain amount of water between 6 and 13 H. After 13 H, a second phase of water arrives in the center of the starch-glycerol blend. There is a decrease of C3 between 16 H and 22 H immersion which may be due to the continuous release of glycerol and/ starch fragments outside the sample.

At 1.72 mm, C1 shows a first "wave" of water that arrives at around 1.30 H followed by a more progressive wave from 3.30 H of imbibition. These hydration steps arrive faster than at the center of the cylinder (at 0 mm) since the profiles corresponds to the interior of the sample. C2 shows an increase in signal up to 4 H and then a decrease. This is the second phase of water that goes out due to strong interactions with starch and collapse of the glycerol cavities. The C3 component increases until 2 H and then slightly decreases by keeping the half of its initial intensity since it corresponds to the water and glycerol in and outside the sample, respectively. The 2.31 mm distance from the center of the sample corresponds geometrically to the edges of the starch-glycerol blend at the beginning of imbibition. Then, as the sample swells during hydration, this position is inside the extrudate. At this position, the C1 component is the highest intensity component assigned to water in the starch-glycerol blend. Its intensity increases due to the entry of water in the sample up to 4 H then the intensity decreases (fixing of the water to starch, the contrast decreases) up to 13 H and then increases again (new water phase). C3 increases rapidly and then more gradually until 13 H to decrease very slightly due to the glycerol release. The C2 component decreases very quickly because at the beginning of the hydration, this position corresponds to a mixture of aqueous phase around the starch-glycerol blend and the edge of it. The latter inflates by absorbing water, which explains that the component C2, characteristic of a water phase relatively mobile sees its signal decrease by strong interaction with the matrix.

Finally, the 3.15 mm distance from the center of the starch-glycerol blend corresponds geometrically to the aqueous phase around the starch-glycerol blend. C2 is the highest intensity component. It is attributed to the aqueous phase around the extruded. Its intensity decreases at the beginning of the imbibition due to the release of glycerol (decrease of contrast because decrease of T2). The C1 signal that corresponds to the water in the starch-glycerol blend is almost null, while C3 increases its intensity gradually with a maximum reached after 13 H when the second hydration phase reaches the center of the starch-glycerol blend.

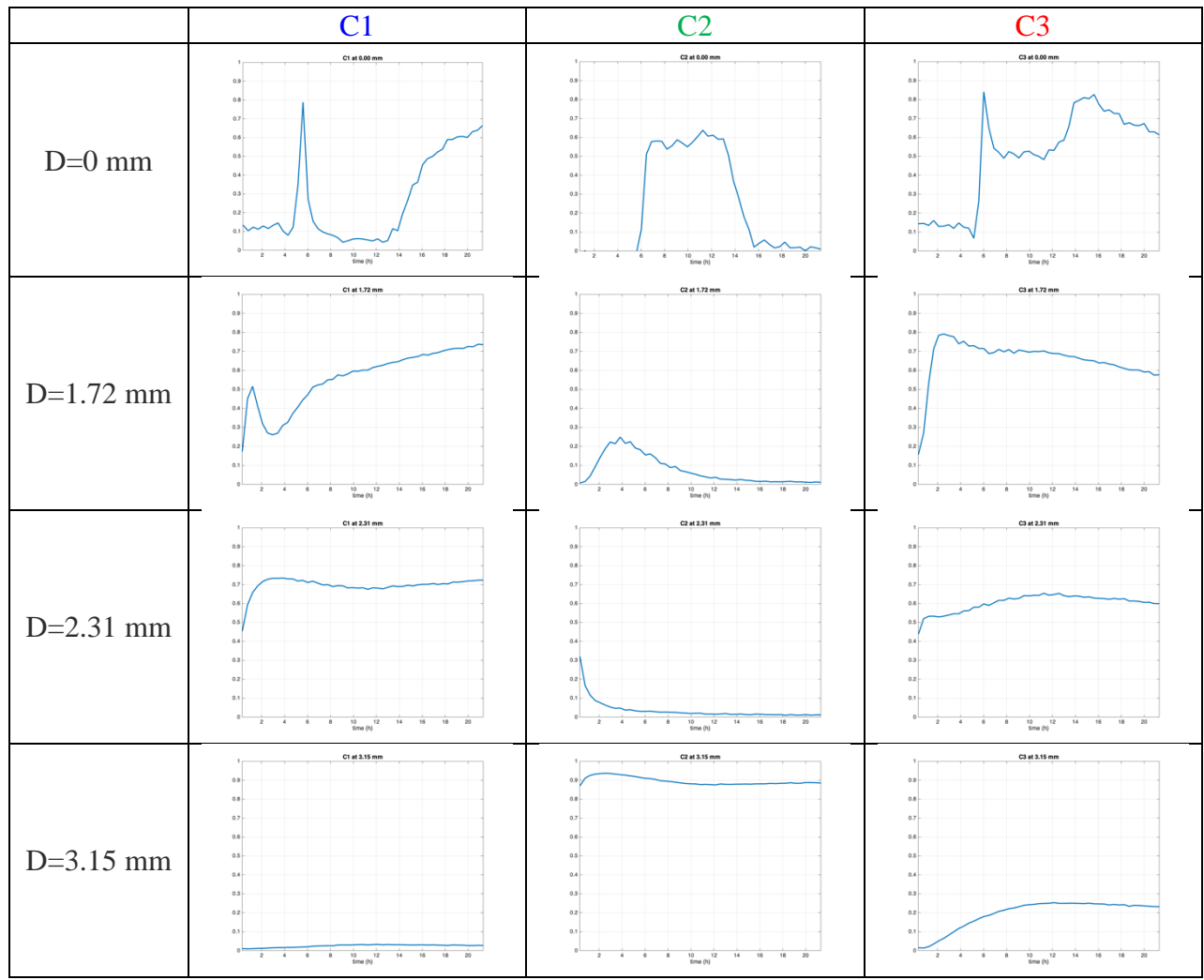

Figure 6. Concentration evolution of the 3 species involved in the swelling process of starch-glycerol blend (C1, C2 and C3), at four distances from the center: 0 mm, 1.72 mm, 2.31mm, 3.15 mm.

In order to relate water transfer to morphology transformation, microscopy experiment have been performed. The figure 7 shows SEM images of sample surface cuts at different times of immersion in water corresponding to the kinetics given by the C1 signal (right part of the figure). The surface of the dry sample seems homogeneous without noticeable asperity or roughness (not shown). After 0.78 H of imbibition, in the domain corresponding to the first and second hydration front, we observe egg-shaped objects of about 100 µm size. They are non-transformed starch granules embedded into the amorphous matrix as shown in previous work [27]. It corresponds to the initial composite structure losing its cohesion during sample swelling and relaxing. Simultaneously, swelling induces cracks and elongated macropores at the granules-matrix interface. The center part of the cylinder not still hydrated, remains smooth and homogeneous. After 2.52 H of imbibition, the domain of « swelled structures » progresses according to the image of the kinetic profile. The microscopy image obtained after 21.30 H of imbibition shows a complete swelled and relaxed structure. Despite microscopy does not evidence two fronts of hydration, we can put forward two hypotheses on mechanisms: (i) the first front corresponds to rapid water penetration in large pores giving a NMR signal that evolves rapidly into the extrudate. During this step, there is a rapid exchange of glycerol by water, confirmed by the presence of glycerol detected in the surrounding environment. After this first front, the water progress into the extrudate slows down due to high water-starch interactions (hydrogen bonds). The second front corresponds to slower water penetration in smaller pores by taking into account these strong water-starch interactions in the relaxed structure. (ii) The second hypothesis is a rapid water penetration by the continuous starch matrix (first front) followed by a slower water penetration inside the semi-crystalline starch granules embedded into the amorphous matrix (second front).

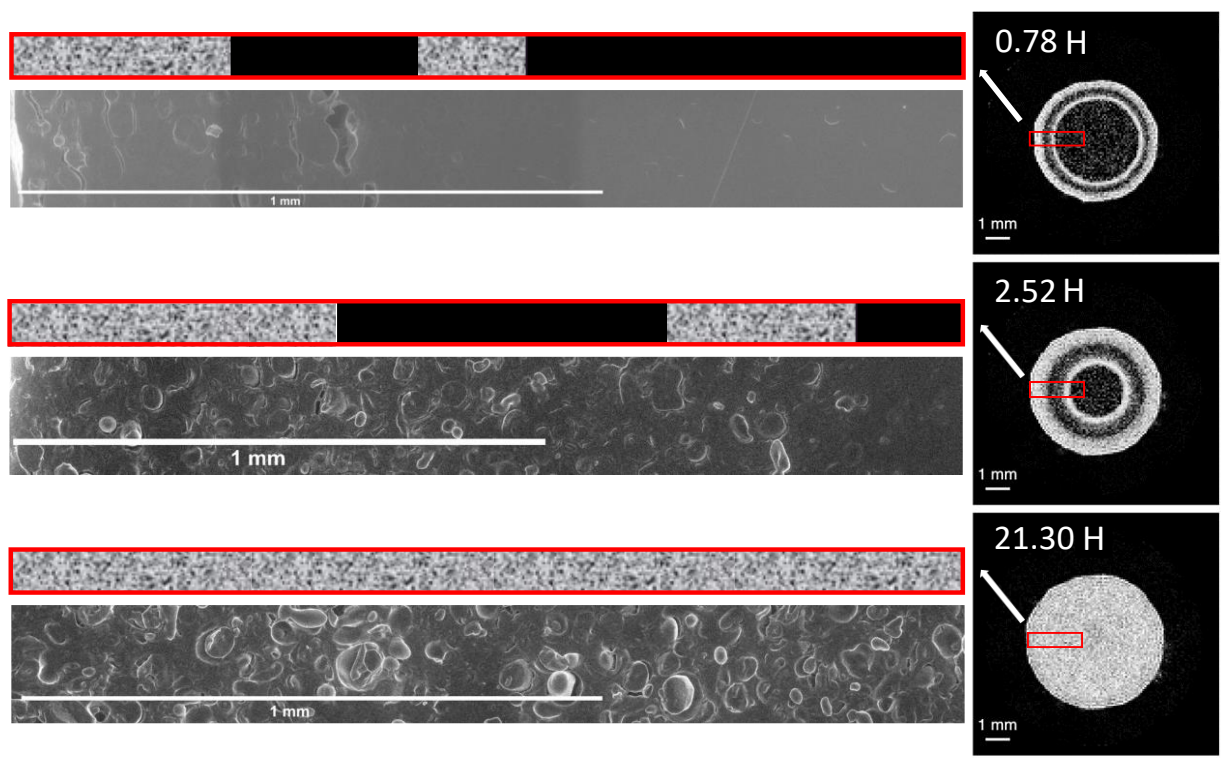

Figure 7. SEM images of sample surface cuts at 0.78 H, 2.52 H, 21.30 H corresponding to the kinetic given by the C1 signal (right part of the figure).

Similar kinetics and spectral $T_2$ values to those obtained using conventional analysis [19,26] were obtained using the proposed methodology. Now, the resolved MCR resolved spectra, which are noise-filtered, and compound-specific image information, can be processed by an ILT without numerical instability problems. Besides, this methodology provides the distribution maps and their associated videos of the evolution of the compounds involved in process that allows interpreting and predicting the physical/chemical phenomena of the system. Therefore, the coupling of MRI measurements and MCR-ALS seems a real promising approach that can be applied to many systems as proteins, microorganisms' media, foodstuffs, any samples containing water, with the aim of understanding and measuring the water ingress in native or processed samples.

Further studies are under progress to process MR image series using their third dimension. The acquisition of successive slices would induce spatial constraints that could be introduced in MCR-ALS analyses. This innovative approach would make it possible to study the homogeneity of the sample over the three dimensions of images. Other constraints as physico-chemical or biological models of multi-scale water transfers could also be potentially used in MCR-ALS analyses in order to improve the processing of MRI images.

## 4. Conclusions

Combination of MRμI and MCR-ALS has proven to be an excellent methodology for investigating water transfers in bio-based products using a non-destructive and non-invasive approach. Particularly, in this study, the application of MCR-ALS multiset analysis on a series of images over the water uptake of a starch-glycerol extruded blend has provided a good description of the various water transfers monitored from a global (image) and a local (pixel) point of view.

Three signal components were resolved by MCR-ALS to describe the system. One corresponded to doped water imbibing the starch-glycerol blend, which decreases thought along the swelling process. The two other components could be assigned to two distinct processes of water transfer of the starch-glycerol blend, which increased during the imbibition. The concentration evolution of these three species at four positions into the sample clearly indicated two times at 6 H and 13 H when two different waterfronts reached the center of the starch-glycerol extruded blend. The resolution of two waterfronts into the blend could be identified thanks to the MCR-ALS method that allows the detection of very weak signals, which are very difficult to resolve by any other mathematical method. Complementary SEM observations permitted to propose two hypotheses of interpretation of these two waterfronts.

The strategy proposed opens a new way to model water transfers monitored by magnetic resonance imaging, without the use of a physical model describing the signal evolution, without application of controversial and time-consuming processing techniques, known to be ill-conditioned and ill-posed problems and using the spatial information contained in the whole image or part of it.

## 5. Acknowledgements

This research did not receive any specific grant from funding agencies in the public, commercial, or not-for-profit sectors.

The authors kindly acknowledge Bruno Novales (BIBS Unité BIA_INRA, Biopolymères, Interactions et Biologie structurale) for SEM experiments.

Magnetic resonance microimaging has been performed using the PRISM core facility (Biogenouest, Univ. Rennes, Univ Angers, INRAE, CNRS, France).

## 6. Conflict of interest

The authors report there are not conflicts of interest

Movie S1. The full set of MCR distribution maps of the three species involved in the swelling process of starch-glycerol blend displaying their evolution process from 0.31H to 21.30H.
.